\documentclass[prl,twocolumn,showpacs,preprintnumbers,amsmath,amssymb]{revtex4}
\usepackage{dcolumn}
\usepackage{bm,amsmath,amssymb}
\usepackage{graphicx}
\usepackage{bm}
\begin{document}
\title{Recrystallization of glass: homogeneous vs. heterogeneous nucleation in La$_{0.5}$Ca$_{0.5}$MnO$_3$.}
\author{A. Banerjee, Kranti Kumar and P. Chaddah}
\affiliation{UGC-DAE Consortium for Scientific Research (CSR),\\ University Campus, Khandwa Road, Indore, 452 017, INDIA}
\date{\today}
\begin{abstract}
We probe through magnetization and resistivity measurements a kinetically arrested glass-like but long-range ordered magnetic state. The transformation kinetics of the magnetic field-temperature induced broad first-order transition from ferromagnetic-metallic (FMM) to antiferromagnetic-insulating (AFI) state gets hindered at low temperature in a La$_{0.5}$Ca$_{0.5}$MnO$_3$ sample. A fraction of high-temperature FMM phase persists to the lowest temperature, albeit as a non-ergodic state. We present a phenomenology for this glass-like but long-range order FMM phase which devitrifies on heating and converts to equilibrium AFI phase. The residual kinetically arrested FMM phase can be `recrystallized' to AFI state by annealing and more efficiently by successive annealing, presumably by heterogeneous nucleation. This glass-like state shows a stimulating feature that when the fraction of glass is larger the `recrystallization' is easier. 
\end{abstract}
\maketitle
Although glass-like behavior is observed in variety of systems including vortex matter \cite{Fish} or monatomic metal \cite{Bhat}, a quantitative understanding of glasses remains a major scientific challenge \cite{Greer, Braw, Deb}. Glass-like non-ergodic state in spin-glass is distinct from other magnetic systems showing metastable magnetism like superparamagnet\cite{Binder}. The high-temperature spin configuration (disorder) is frozen in time below a certain temperature in spin-glass, despite structural order, resulting in a new paradigm for glassy system. Yet, new vista of glassy magnetic behaviour is revealed where the high-temperature long-range structural as well as spin orders get arrested below a certain temperature (T$_K$) and glass-like non-ergodic state is created in systems ranging from intermetallics \cite{Chat, Sampath, Roy1}, shape-memory alloys \cite{Sharma}, magnetocaloric materials \cite{Roy2}, colossal magnetoresistance (CMR) manganites \cite{Kumar, Wu, Ban1, Ban2, Rawat} etc. Similar to the structural glass, in these systems also, kinetics of the first-order transformation is arrested while preserving the high-temperature phase, albeit as a non-ergodic state. The first-order transition temperature (T$_C$) in real systems will have a distribution, because of the intrinsic disorder, for regions having dimension of the order of correlation length\cite{Imry}. Consequently, the superheating (T**) and the supercooling (T*) limits will also have a distribution resulting in a broad hysteretic transition. On similar account, it has been observed that, the first-order field (H) and temperature (T) induced transitions are broadened with respect to the control variables like H and T \cite{Soibel, Roy3}. This naturally implies that the arrest of kinetics leading to formation of glassy phase for the respective regions would also occur over a broader section in the control variable space (like H-T space) \cite{Man, Chaddah}. Arrest of kinetics in H and T induced broad first-order transformation process connecting two phases with long-range structural and also magnetic order results in persistence of high-T phase at low-T, where it is energetically unstable and coexists with the low-T equilibrium phase \cite{Chaddah}. Similar to the conventional glass, where the glass formation can be tuned by control variables like pressure (P) and T \cite{Mis}, in case of magnetic transitions H and T plays the equivalent role \cite{Kumar, Wu, Ban1, Ban2, Rawat}.
  
Devitrification is an evidence of glassy state \cite{Greer, Braw, Deb} which is recently observed in a variety of such magnetic cases \cite{Roy2, Ban1, Ban2}. Here we present some intriguing features of a glassy system using a manganite around half doping, viz. La$_{0.5}$Ca$_{0.5}$MnO$_3$ (LCMO). This is similar to the sample where Loudon et al. have shown a first-order transformation from high-T ferromagnetic to low-T antiferromagnetic phase \cite{Loud}, with a coexistence of ferromagnetic regions at low-T, below the closure of hysteresis. We show that the AFI state is in equilibrium at low-T and the FMM phase fraction exists as kinetically arrested glassy or non-ergodic state, based on magnetization (M) and resistivity (R) measurements similar to those reported earlier \cite{Ban2, Rawat}. Further, on annealing, this glassy FMM phase fraction `crystallizes' to equilibrium AFI phase similar to the formation of nano-crystals within metallic glasses or in glass ceramics \cite{Greer}. Successive annealing can convert larger fraction of glass to the `crystalline' state; the glassy FMM phase of LCMO also shows similar additional conversion to AFI state. The most significant as well as intriguing result of this study is the observation that even much larger fraction of glass can be `recrystallized' when the starting glass fraction is larger for both single and successive annealing. This implies that `crystallization' from glass is more efficient when the starting fraction of glass is larger.
 
Polycrystalline LCMO sample has been prepared through a chemical route known as `pyrophoric method'. High purity ($\geq99.9\%$) La$_2$O$_3$, CaCO$_3$ and C$_4$H$_6$MnO$_4$.4H$_2$O are taken in stoichiometric quantities as starting materials. These materials are dissolved in aqueous nitric acid and the resulting solutions are mixed together with triethanolamine (TEA). The complex solution is heated to dehydrate and decompose leaving behind organic-based, black, fluffy precursor powder. This dried mass is then ground to fine powder, palletised and then calcined at $1000^{o}C$ for 3 hours in oxygen atmosphere. The powder x-ray diffraction (XRD) is carried out using an 18 kW Rigaku Rotaflex RTC 300 RC diffractometer with $CuK_\alpha$ radiation. Rietveld profile refinement of XRD pattern confirms that the sample is in single phase without any detectable impurity and crystallizes in orthorhombic structure with `pnma' space group. It is known that by varying preparation condition, different samples with nominal composition of La$_{0.5}$Ca$_{0.5}$MnO$_3$ can be prepared having different fraction of FMM phase at low temperature \cite{Levy}. For this study we choose one which is close to that used by Loudon et al. \cite{Loud}, however, we would like to emphasise that the exact details of the sample may not be the most important factor for the present study. The resistivity and magnetic measurements are performed using commercial set-ups (14Tesla-PPMS-VSM, M/s. Quantum Design, USA).

Non-ergodicity in the sample is demonstrated in Fig.1. Each time, the sample is cooled from 320K to 5K in different H and then M is measured while changing the field isothermally from H to -H (Fig.1(a)). Higher cooling field renders larger fraction of ferromagnetic phase which remains frozen at 5K even when the H is reduced to zero as shown by distinct magnetization in the -ve field cycles, which is a mirror image of the initial field cycle. When the same measurement is repeated at 25K, as shown in Fig. 1(b), there is partial devitrification of the glassy FMM phase while reducing the field to zero. This is evident from the -ve field cycles showing lower magnetization than the starting values. The reduction is larger for the higher cooling field because of the nature of the kinetic arrest band as discussed earlier \cite{Kumar, Chaddah}, which we briefly touch upon in the next paragraph. Thus Fig. 1(a) shows that, cooling in different fields can control the fraction of glass-like arrested state, moreover, this fraction remains unchanged when the field is withdrawn at 5K. The nature of this arrested phase is probed through M-T measurements in 1 Tesla field following different protocols as shown in Fig. 1(c). The disorder-broadened first-order transition from FMM to AFI phase with reducing temperature is evident from the hysteresis between field-cooled cooling (FCC) and field-cooled warming (FCW) paths. However, similar to the Ref. \cite{Loud}, a sizable magnetization at 5K, much below the closure of hysteresis, suggests the persistence of a fraction of ferromagnetic phase. Even after zero-field cooling (ZFC) there is substantial magnetization. This clearly indicates that the FMM to AFI transition is arrested and remains incomplete (even in zero field) at low-T.  After cooling the sample in 4T and 6T, H is isothermally changed to the measurement H of 1T at 5K in both cases. Significantly larger but different magnetizations are now observed at the same H and T (1T and 5K) indicating non-ergodicity arising from additional glass-like arrested FMM phase. On warming, these states devitrify and the system approaches the equilibrium AFI phase as shown by rapid fall in M. Figure 1(d) shows similar measurement in 3T after cooling in different H including zero field along with the FCC and FCW branches for 3T. It may be noted that, the magnetization evolve differently for the cases where cooling H are larger than the measuring H from the cases where cooling H is smaller than the measuring H. While the larger fraction of FMM phase accrued during cooling in H higher than the measurement H devitrify on heating, no devitrification is evident in the presence of measurement H if the cooling fields are smaller. This also indicates that the low-T equilibrium state is AFI akin to the recent observations made in other half-doped manganites \cite{Ban2, Rawat}.

To acquire semiquantitative understanding about the observations of Fig.1 from the heuristic H-T phase diagram proposed for the similar systems \cite{Chaddah}, we redraw it for the present case in Fig. 2. The supercooling (or superheating) line in the H-T space, across which the transformation takes place is broadened by disorder into a band formed by a quasi-continuum of lines representing various regions of the sample. Sign of its slope reflects that the FMM phase will exist over a larger temperature range in higher fields. While cooling, if the system crosses this band unhindered, then it will covert to equilibrium AFI phase at low temperature. However, in the present system, the glass-like arrest of  the transformation is depicted by a quasi-continuum of lines, forming (H$_K$, T$_K$) band. Sign of the slope of (H$_K$, T$_K$) band is consistent with the observation that when increase in temperature causes devitrification to AFI state (Fig. 1(c) and 1(d)), then isothermal devitrification must be observed on lowering the field as shown in Fig. 1(b). As the sample is cooled in zero H towards point A, we have a homogeneous FMM phase even though it is metastable. At point A, the entire sample is above the supercooling spinodal and no homogeneous nucleation takes place. Since regions corresponding to W-band get kinetically arrested at point B, before there is any homogeneous nucleation, these regions remain FMM at the lowest temperature of point C. But regions corresponding to bands X,Y and Z have got converted to AFI at this point C since their respective spinodal T* is higher than their corresponding T$_K$. We now warm the sample to point B. We retain X, Y and Z in AFI phase, but W is kinetically arrested. We now warm toward point A. W-band is no longer kinetically arrested, is above its supercooling spinodal, but is sitting in an environ where many nuclei of AFI phase exist corresponding to bands X, Y and Z. There is now a possibility of  FMM phase of band W undergoing heterogeneous nucleation and converting to the AFI phase. A closer scrutiny of the data on the present system will also justify an anticorrelation between the  (H*, T*) and (H$_K$, T$_K$) band, which was proposed earlier \cite{Chaddah} and experimentally verified for a variety of systems \cite{Kumar, Roy2, Rawat}. 

Now we show in figure 3, the process of `recrystallization' of the glassy FMM phase through the magnetization measurement. Fig. 3(a) shows M-T in 1T while warming from 5K. After cooling in 1T, the FCW path starts with a magnetization value of about 0.71 $\mu$$_B$/Mn at 5K, whereas after cooling in zero H the ZFC path has a starting value of about 0.6 $\mu$$_B$/Mn. However, much smaller M of about 0.5 $\mu$$_B$/Mn can be achieved when the sample is cooled to 5K in 1T and then warmed to 130K and cooled back to 5K. While warming, this 130K annealed state evolve distinctly from the FCW or ZFC paths and finally merge with them on approaching the superheating limit around 200K. The effect of successive annealing is shown in Fig. 3(b) through the M-H measurements at 5K after initially cooling in 2T from 320K. After initial cooling the field is cycled from 2T to -2T at 5K. Then the sample is warmed to 60K and cooled back to 5K where a M-H is measured between $\pm$ 2T. Annealing at 60K has not only substantially reduced the glassy FMM phase fraction, the magnetization value in 1T is even less than the ZCF value of 0.6 $\mu$$_B$/Mn at 5K in 1T. The sample is then successively annealed at higher temperatures and each time M-H is measured after cooling to 5K showing lower fraction of FMM phase. This way, after annealing to 130K the magnetization at 5K in 1T is reduced to about 0.43 $\mu$$_B$/Mn which is lower than 0.5 $\mu$$_B$/Mn achieved after single annealing at 130K (Fig.3(a)). It is proposed earlier that the AFI phase `crystallizes' from high temperature FMM phase through the `nucleation and growth' process \cite{Loud}. During the initial cooling, the `homogeneous nucleation' process is dominant because there is no AFI seed in the sample \cite{Deb2}. During annealing the existing AFI phase provides seeds for heterogeneous nucleation leading to additional `recrystallization'. It is possible to achieve even more `recrystallization' of the residual glassy FMM phase to AFI state by successive annealing \cite{Greer}. It may be noted here that annealing in temperatures higher than 150K results in opposite effect because of back conversion to FMM phase.

One intriguing feature of this glassy magnetic state is that, it is easier to `recrystallize' the glass when its starting fraction is larger. It is already shown in Fig. 3(a) that annealing the 1T cooled state to 130K produces lower magnetization than the zero-field cooled state. Figure 4 shows that, larger fraction converts to AFI phase when the starting FMM fraction is larger at 5K.  It is shown in Fig. 1(c) that larger fraction of FMM phase can be created at 5K if cooled in 6T, and Fig. 4(a) shows that if this 6T cooled state is annealed to 130K then it produces larger fraction of AFI phase and lower magnetization in 1T at 5K, which is even lower than the 1T cooled state annealed to 130K. Yet, larger conversion to AFI state can be achieved when the 6T cooled state is subjected to successive annealing at progressively higher temperature. After cooling in 6T to 5K the field is reduced to zero and the sample is subjected to a temperature cycling between 5K and T$_H$, where T$_H$ starts from 60K and each time it is increased by 10K to go up to 130K. Much lower magnetization is measured at 5K in 1T after this temperature cycle, which evolve distinctly on warming and merges with all others on approaching the superheating spinodal. This half-doped manganite is expected to have the spin-aligned value of 3.5 $\mu$$_B$/Mn accompanied by metallic conductivity \cite{Loud}. The observed variation in magnetization at 5K from 0.6 $\mu$$_B$/Mn to 0.37 $\mu$$_B$/Mn can be attributed to a variation of frozen FMM phase fraction from about 17\% to 11\%, which is close to the percolation threshold for electrical conductivity. Hence, around this FMM phase fraction resistivity changes can be more drastic which is vividly demonstrated in Fig. 4(b) by colossal change in resistivity measured in zero field while warming from 5K. The lower resistivity measured while warming from 5K after cooling in zero field increases many fold if the sample is annealed at 140K. The resistivity increases further when the sample is cooled to 5K in 1T and then annealed to 140K after reducing the field isothermally to zero at 5K. Yet, orders of magnitude increase in resistivity is observed after cooling in 6T and then reducing the field isothermally to zero at 5K followed by an annealing at 140K. The increase in resistivity is so gigantic that it exceeds the measurement range of the instrument until the sample is heated above 38K. Successive annealing produces much larger resistivity for each case (not shown in the figure for clarity). Contrary to earlier report of thermal cycling on a manganite sample \cite{Mahe},  the observed changes are reversible and can be reproduced once the superheating limit is crossed or the sample is heated to about 250K. There is no permanent change in the sample which is also evident from the fact that, the successive annealing increases the resistivity, at 5K, up to the annealing temperature of 150K and above that it decreases \cite{Ban3}. In this case, the experiments are designed to probe different aspects of the glassy magnetic state, which can be created or annihilated in the same sample by following different H-T paths below room temperature. \emph{The observation of better conversion to `crystalline' state when the starting fraction of glass is more is rather intriguing} and may have some connection to the counterintutive observation that ``hot water can freeze faster than cold" \cite{Jeng}.
   
In conclusion, it is shown that the FMM long-range ordered magnetic state behaves like a glass at low-T in the LCMO sample even when cooled in zero field. This state devitrifies on heating, akin to canonical glass, to equilibrium AFI phase. This glassy state `recrystallize' to AFI state on annealing and more easily by successive annealing presumably through heterogeneous nucleation process, producing gigantic change in resistivity. In this case the glass formation can be easily tuned by H and T compared to tuning it by P and T for conventional glass. \textit{Larger fractions of glass `recrystallize' more easily.} Such intriguing features of this glassy system can be probed even in zero field by microscopic tools, mapping the change in phase fractions using techniques like photoemission, MFM, STM etc. This should shed more light on the physics of glasses.

DST, Government of India is acknowledged for funding the 14T-PPMS-VSM at CSR, Indore.

\newpage
\begin{figure*}
	\centering
		\includegraphics{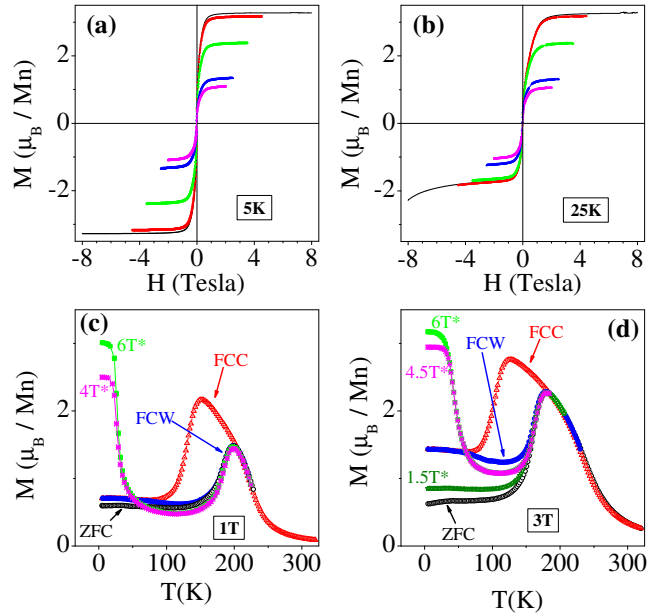}
	\caption{(color online). Magnetization of La$_{0.5}$Ca$_{0.5}$MnO$_3$. (a) The sample is cooled from 320K in different H= 2, 2.5, 3.5, 4.5 and 8 Tesla and M vs H is measured at 5K while isothermally changing the H from +H to -H. (b) M vs. H at 25K after following the same cooling protocol as panel (a). (c) M vs. T measured in 1T field. FCC and FCW magnetizations are measured while cooling and warming in 1T. For ZFC path, the sample is cooled in zero field. Again the sample is cooled from 320K in 4T or 6T and the field is isothermally changed to 1T at 5K for M-T while warming and marked in the figure as 4T* and 6T* respectively. (d) M-T in 3T. FCC and FCW are measured while cooling and warming in 3T, ZFC is after cooling in zero field. Again the sample is cooled from 320K in 1.5, 4.5 and 6T and the field is isothermally changed to 3T for M-T while warming and marked in the figure as 1.5T*, 4.5T* and 6T* respectively.    }
	\label{fig:FIG1}
\end{figure*}
\newpage
\begin{figure*}
	\centering
		\includegraphics{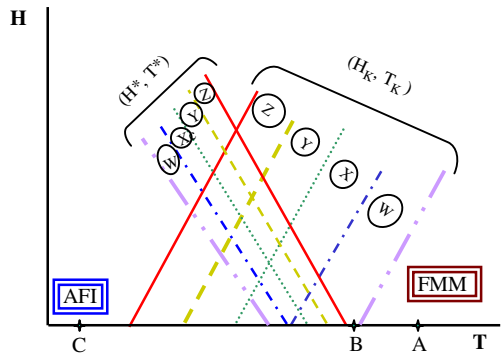}
	\caption{(color online). The heuristic H-T diagram for La$_{0.5}$Ca$_{0.5}$MnO$_3$. The high-T phase is FMM and the low-T phase is AFI. The FMM phase can be supercooled to the (H*, T*) band. To represent broad H-T induced first order transition, only 4-regions (W, X, Y and Z) of the sample are shown in this schematic. The glass-like arrest of kinetics will occur at (H$_K$, T$_K$) band. Anticorrelation between supercooling and kinetic arrest is assumed (see ref. \cite{Roy2, Kumar, Chaddah} for details). Points A, B and C are as referred to in text.}
	\label{fig:FIG2}
\end{figure*}
\begin{figure*}
	\centering
		\includegraphics{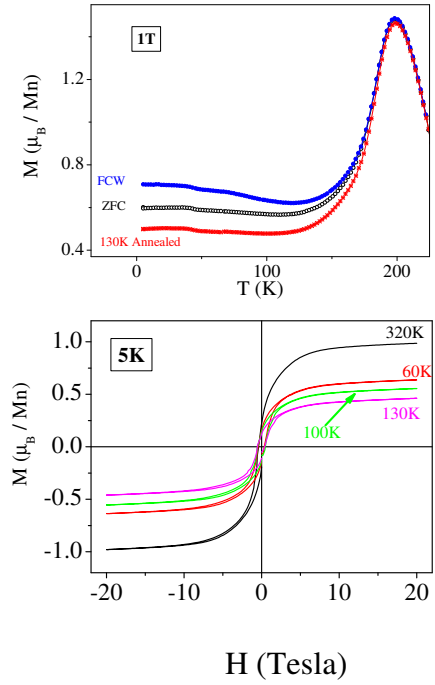}
	\caption{(color online). Effect of annealing and successive annealing on the magnetization of La$_{0.5}$Ca$_{0.5}$MnO$_3$. (a) M-T while warming in 1T after cooling in zero field and 1T. Again after cooling in 1T the sample is annealed at 130K and M-T is measured after cooling back to 5K. (b) After cooling the sample in 2T from 320K to 5K the M-H is measured cycling the field between +2T to -2T. Then the sample is annealed at progressively higher temperatures and M-H is measured at 5K after cooling back from each annealing temperature as marked in the figure.}
	\label{fig:FIG3}
\end{figure*}
\begin{figure*}
	\centering
		\includegraphics{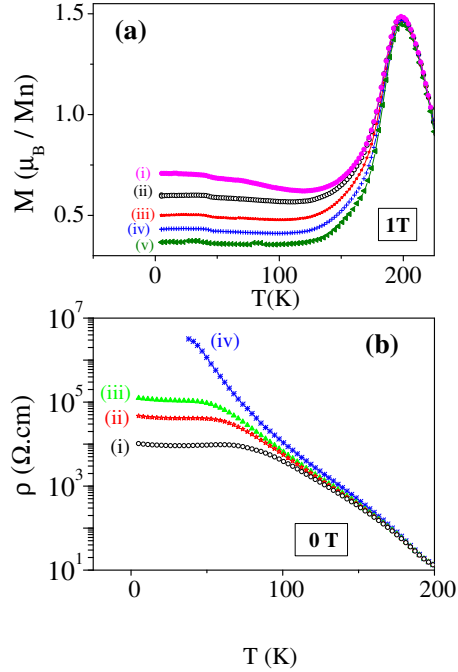}
	\caption{(color online). Effect of starting glass fraction on the `recrystallization' of AFI phase in La$_{0.5}$Ca$_{0.5}$MnO$_3$. (a) M-T while warming in 1T. Sample is prepared at 5K through (i) FCC in 1T, (ii) ZFC in 1T, (iii) Annealed at 130K after FCC in 1T, (iv) FCC 6T and annealed once at 130K in zero field, (v) FCC 6T and annealed successively to 130K as described in text. (b) R-T in zero field while warming. Sample is prepared at 5K through (i) ZFC, (ii) ZFC and annealed at 140K, (iii) annealed at 140K after FCC in 1T, (iv) annealed at 140K after FCC in 6T.}
	\label{fig:FIG4}
\end{figure*}
\end{document}